\documentclass[pre,twocolumn,preprintnumbers,showpacs,floatfix]{revtex4}
\usepackage{graphicx} 
\usepackage[toc,page]{appendix}
\usepackage{epsfig}
\usepackage{epstopdf}
\usepackage{pslatex}
\usepackage{subfig}
\usepackage{amsmath}
\usepackage{amssymb}
\usepackage{multirow}
\usepackage{color,soul}
\begin{document}
\title{Percolation and jamming transitions in particulate systems with and without cohesion\footnote{To appear in Phys. Rev. E.}}
\author{L. Kovalcinova, A. Goullet, L. Kondic}
\affiliation{
Department of Mathematical Sciences,
New Jersey Institute of Technology,
University Heights,
Newark, NJ 07102
}

\date{\today}

\begin{abstract}
 We consider percolation and jamming transitions for particulate systems exposed to compression.    For the
systems built of particles interacting by purely repulsive forces in addition to friction and viscous damping, it is found that 
these transitions are influenced by a number of effects, and in particular by the compression rate.   
In a quasi-static limit, we find that for the considered type of interaction between the particles, percolation and jamming transitions 
coincide.  For cohesive systems, however, or for any system exposed to even slow dynamics, the differences
between the considered transitions are found and quantified.
\end{abstract}


\maketitle

\section{Introduction}
The dense systems of particles interacting by either purely repulsive potentials, such as dry granular particles, or 
by both repulsive and attractive ones, such as wet granulates, appear virtually everywhere, from nature to a variety 
of applications bridging the scales from nano to macro.   The structure of the force field by which the particles interact
may be very complex, in particular on meso-scales where this force field is nonuniform and forms force networks.   
These networks are of relevance not only to granular systems, but to many other ones, such as foams and colloids.  
Their properties  have been recently explored using a variety of different approaches, ranging from theoretical and 
computational ones based on exploring local structure of force networks~\cite{peters05}, networks type of approaches~\cite{daniels_pre12,herrera_pre11}, and
topological methods~\cite{epl12,pre13,arevalo_pre13}.

While percolation has been considered for dense particulate systems~\cite{aharonov99,arevalo_pre10,lois,ohern_pre_12}, much more is known about 
static and ordered lattice-based systems~\cite{stauffer,albert_barabasi_02}, for which two types of percolation are 
discussed -- rigidity and connectivity percolation~\cite{alexander_physrep05, guyon90}.   However,
lattice models do not account for nonlinear effects at particle contacts, such as friction and 
viscous damping, or for dynamics, so it is unclear whether the results obtained for lattice systems apply to particulate ones~\cite{guyon90}. 
For the latter, the connection between 
percolation (connectivity) and jamming (rigidity) transitions was discussed recently for both non-cohesive and cohesive frictionless systems, 
and it was found (for the systems considered) that these two transitions in general differ~\cite{lois,ohern_pre_12}.  However, these conclusions
were reached by considering rather specific interaction models (over-damped dynamics), and the question whether they hold in general,
and whether they also follow from the models commonly used to simulate physical granular particles, is still open.

In this paper, we discuss the relation between percolation and jamming for frictional and frictionless particles in two 
spatial dimensions, both with and without cohesion.
We consider slowly compressed systems that go through percolation and jamming and discuss how these transitions depend on the system properties.
The motivation for considering compression is that it is a simple protocol that avoids the complexities associated with shear, and allow us to 
focus the discussion.  However, consideration of any dynamics, including compression, naturally leads to the questions related to the rate-dependence 
of the results,  and, as we will see, to new insight  into percolation and jamming transitions for evolving particulate systems.    

The paper is organized as follows.  In Sec.~\ref{sec:sim} we present the simulation techniques.   In Sec.~\ref{sec:results} we present out 
findings, first for purely repulsive systems in Sec.~\ref{sec:repulsive}, and then for cohesive ones in Sec.~\ref{sec:cohesive}.  Section~\ref{sec:conclusions} is devoted to summary, conclusions, and future outlook.  

\section{Simulations} 
\label{sec:sim}

We perform discrete element simulations using a set of circular particles confined in a square domain, using a slow-compression protocol~\cite{epl12,pre13}, augmented
by relaxation as described below. Initially, the system particles are placed on a square lattice and are given random velocities; we have verified that the results are independent of the distribution and magnitude of these initial velocities. The discussion related to possible development of spatial order as the system is compressed can be found
in~\cite{pre13}, and the issue of spatial isotropy of the considered systems is considered later in the text.

In our simulations gravity is not considered, and the diameters of the particles are chosen from a flat distribution of width $r_p$. System particles are soft inelastic disks and interact via normal and tangential forces, including static friction, $\mu$ (as in~\cite{epl12,pre13}). The particle-particle (and
particle-wall) interactions include normal and tangential components.
The normal force between particles $i$ and $j$ is 
\begin{eqnarray}
 {\bf F}_{i,j}^n& =&k_n x {\bf n} - \gamma_n \bar m {\bf v}_{i,j}^n\\
 r_{i,j} = |{\bf r}_{i,j}|, \ \ \ {\bf r}_{i,j} &=& {\bf r}_i - {\bf r}_j, \ \ \  {\bf n} = {\bf r}_{i,j}/r_{i,j}\nonumber
\end{eqnarray}
where ${\bf v}_{i,j}^n$ is the
relative normal velocity.  The amount of compression is $x =
d_{i,j}-r_{i,j}$, where $d_{i,j} = {(d_i + d_j)/2}$, $d_{i}$ and
$d_{j}$ are the diameters of the particles $i$ and $j$. All quantities
are expressed using the average particle diameter, {\bf $d_{ave}$}, as
the lengthscale, the binary particle collision time  $\tau_c = 2\pi
\sqrt{d_{ave}/(2 g k_n)}$ as the time scale, and the average particle mass,
$m$, as the mass scale.  $\bar m$ is the reduced mass, $k_n$ (in units
of ${ m g/d_{ave}}$) is set to a value
corresponding to photoelastic disks~\cite{geng_physicad03}, and
$\gamma_n$ is the damping coefficient~\cite{kondic_99}.  The
parameters entering the linear force model can be connected to
physical properties (Young modulus, Poisson ratio) as described
 e.g. in \cite{kondic_99}.
\vskip -0.02in

We implement the commonly used Cundall-Strack model for static
friction~\cite{cundall79}, where a tangential spring is introduced
between particles for each new contact that forms at time $t=t_0$.
Due to the relative motion of the particles, the spring length,
${\boldsymbol\xi}$ evolves as $\boldsymbol\xi=\int_{t_0}^t {\bf
 v}_{i,j}^t~(t')~dt'$, where ${\bf v}_{i,j}^{t}= {\bf v}_{i,j} - {\bf
 v}_{i,j}^n$.  For long lasting contacts, $\boldsymbol\xi$ may not
remain parallel to the current tangential direction defined by $\bf
{t}={\bf v}_{i,j}^t/|{\bf v}_{i,j}^t|$ (see,
e.g,.~\cite{brendel98}); we therefore define the corrected
$\boldsymbol\xi{^\prime} = \boldsymbol\xi - \bf{n}(\bf{n} \cdot
\boldsymbol\xi)$ and introduce the test force 
\begin{equation}
{\bf F}^{t*} =-k_t\boldsymbol\xi^\prime - \gamma_t \bar m {\bf v}_{i,j}^t
\end{equation}
where $\gamma_t$ is the coefficient of viscous damping in the tangential
direction (with $\gamma_t = {\gamma_n}$).  To ensure that the
magnitude of the tangential force remains below the Coulomb threshold,
we constrain the tangential force to be 
\begin{equation}
 {\bf F}^t = min(\mu |{\bf F}^n|,|{\bf F}^{t*}|){{\bf F}^{t*}/|{\bf F}^{t*}|}
\end{equation}
and redefine
${\boldsymbol\xi}$ if appropriate.

Cohesive forces are modeled using the approach outlined in~\cite{herminghaus}, and are considered to arise from the capillary bridges that form 
when particles get in contact.    The functional form of this force is given by 
\begin{equation}
 F_b={2\pi R\gamma\cos\theta/ (1+1.05\hat s+2.5\hat s^2)}
\end{equation}
where $\hat s=s\sqrt{R/V}$ and $s=r_{ij}-(d_i+d_j)/2$  
(taken to be $\ge 0$) is the particle separation. 
Here, $ {1 /R}=1/2({1 /d_1}+ {1 /d_2})$~\cite{willett_2000} (for simplicity we do not account here for polydispersity and 
use $d_1=d_2=1$ in dimensionless units), and $V$ is
the volume of a capillary bridge between particles. 
In the present work we assume that all capillary bridges are of the same volume. 
 For contact angle, $\theta$, we use  $\theta = 12^\circ$, comparable to the value for (deionized ultra-filtered) water and (clean) glass~\cite{Gu06}.   
For the surface tension, $\gamma$, we use the value corresponding to water, $72$ dyn/cm, scaled appropriately.  The critical separating distance,  $s_c$, at which a 
bridge breaks is given by 
\begin{equation}
s_c = (1+{\theta/ 2})({V^{1/3}/R}+{V^{2/3}/R^2})
\end{equation}
\begin{figure}[t]
\centering
\subfloat[$\bar F = 0$.    ]{\epsfig{file=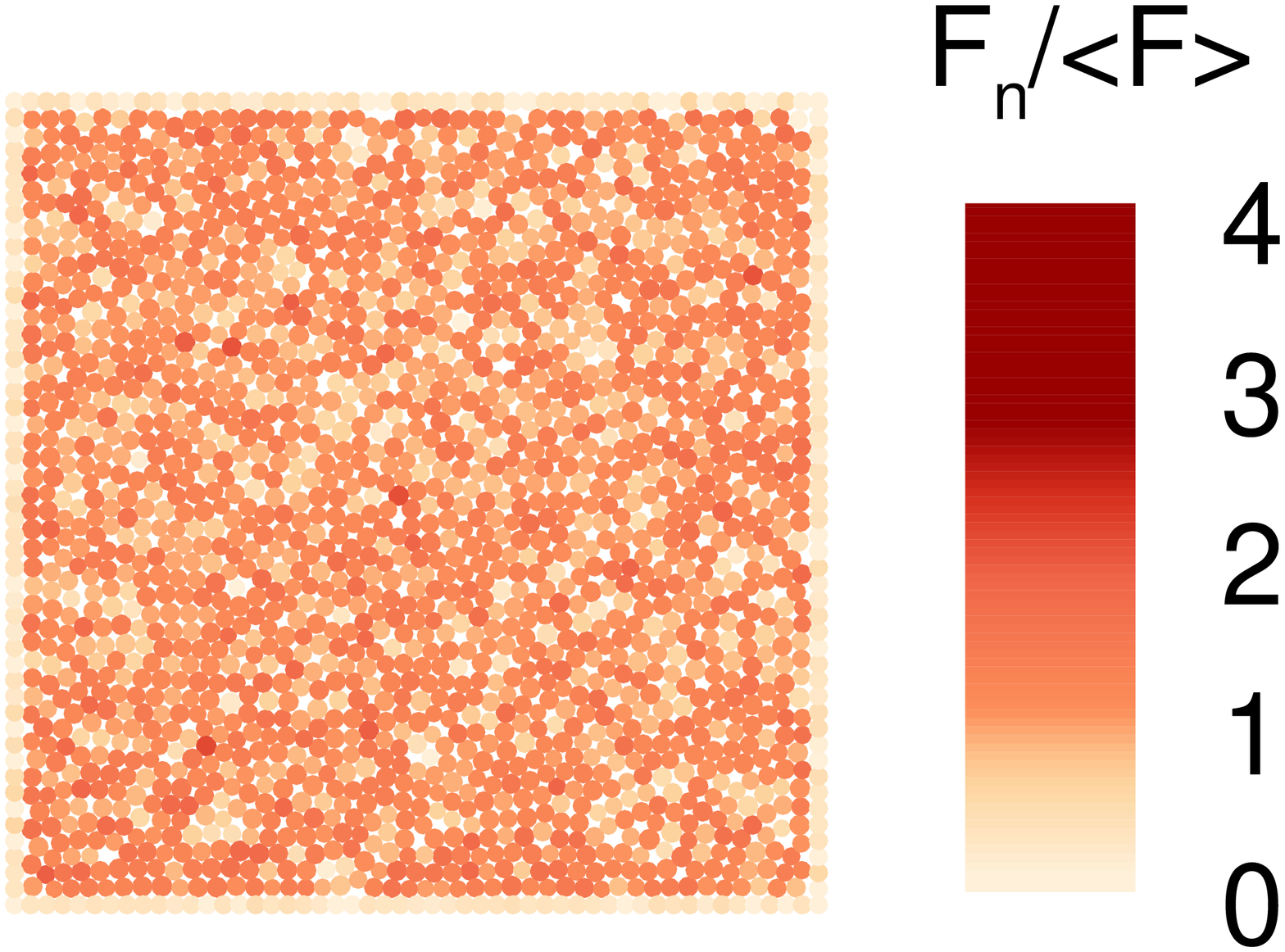,width=1.7in}}
\subfloat[ $\bar F = 1$.   ]{\epsfig{file=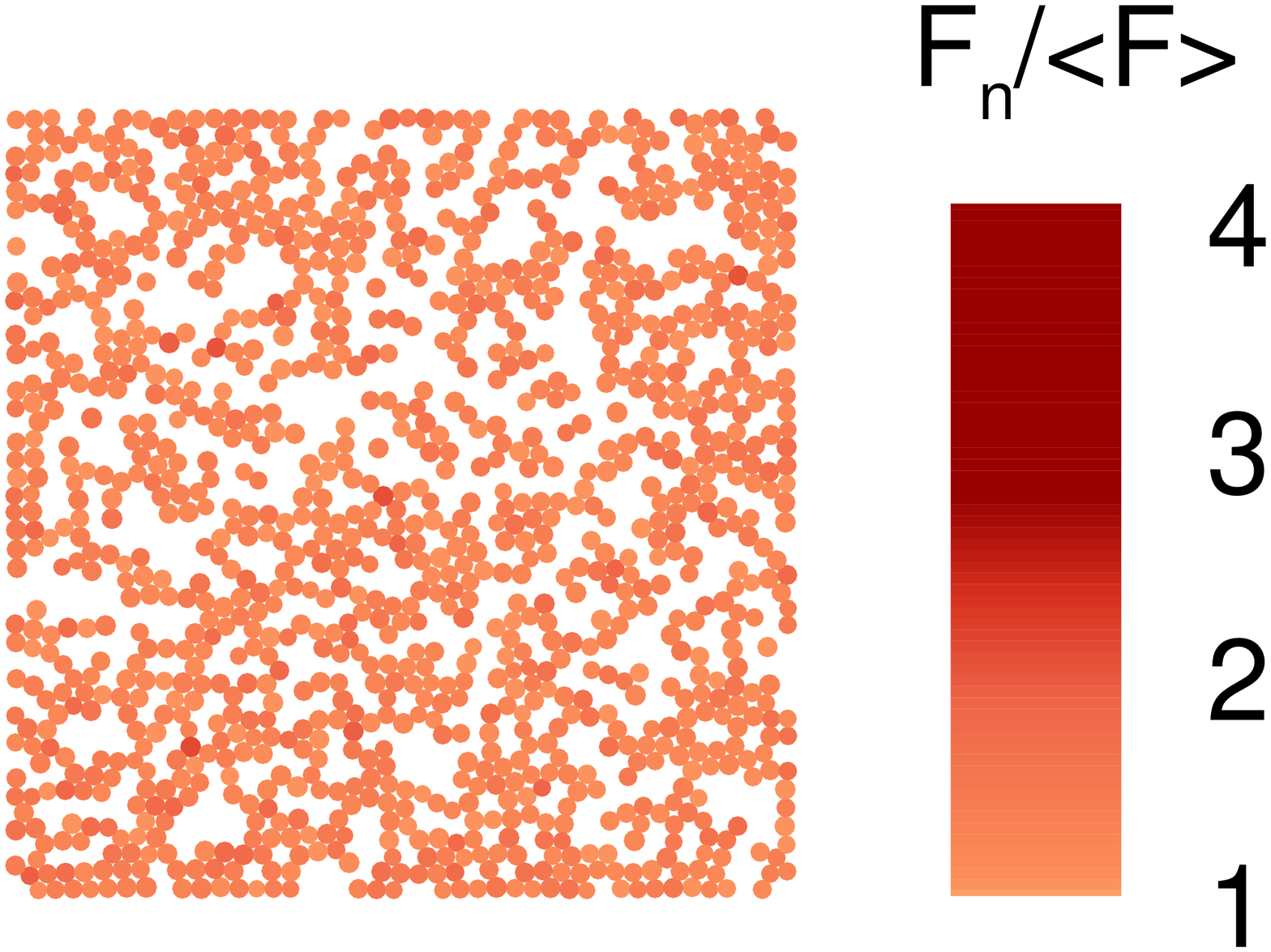,width=1.7in}}
\caption{(Color online) An example of a reference system for different force thresholds at  $\rho=0.9$ (see Supplementary Material in~\cite{epl12} for 
animations).}
\label{part_refcase}
\end{figure}
Here, $s_c$ could be thought of as a measure of the strength of cohesion; larger $s_c$ leads to more pronounced 
cohesive effects.

Our simulations are carried out by slowly compressing the domain,  starting at the packing fraction $0.63$ and ending at 
$0.90$, by the moving walls built of monodisperse particles with diameters of size $d_{ave}$ placed initially at equal distances, $d_{ave}$, from each other. 
The wall particles move at a uniform (small) inward velocity, $v_c$, equal to $v_{0}= 2.5\cdot 10^{-5}$ (in the units of $d_{ave}/\tau_c$),
or a fraction of it, as we explore the influence of compression speed. Due to compression and uniform inward velocity, the wall particles (that do not 
interact with each other)  overlap by a small amount. When the effect of compression rate is explored, $v_c$ is decreased, or the compression 
stopped to allow the system to relax.  In order to obtain statistically relevant results, we simulate a large number of initial configurations (typically $20$),  
and average the results. Due to the compression being slow, we do not observe any different behavior close to the domain boundaries compared to 
the rest of the domain.    

We integrate
Newton's equations of motion for both the translational and rotational
degrees of freedom using a $4$th order predictor-corrector method with
time step $\Delta t =0.02$.    Our reference system is defined by $N=2000$
polydisperse particles ($r_p = 0.2$), with $k_n = 4\cdot 10^3$, $e_n =
0.5$, $\mu = 0.5$, and $k_t = 0.8
k_n$~\cite{goldhirsch_nature05}; the (monodisperse) wall particles have the same physical properties.   Larger domain simulations are carried out
with up to $N =20, 000$ particles. If not specified otherwise, cohesion is not included.
 
\section{Results}
\label{sec:results}

\subsection{Purely repulsive systems}
\label{sec:repulsive}

Figure~\ref{part_refcase}(a) shows an example of the reference system at $\rho=0.90$,  with the particles color-coded according to the total normal force, 
normalized by the average normal force, $F_n/<F>$ (we focus only on the normal forces in the present work). 
If the system contains a set of particles in contact that connects top/bottom or left/right wall, 
then there is contact percolation.   We will also consider force percolation by focusing on the particles sustaining force larger than a
given force threshold and ask how the percolation properties are influenced by a nonvanishing threshold.
 \begin{figure}[ht!]
\subfloat[The percolation probability, $P(\rho,\bar F)$.]{\epsfig{file=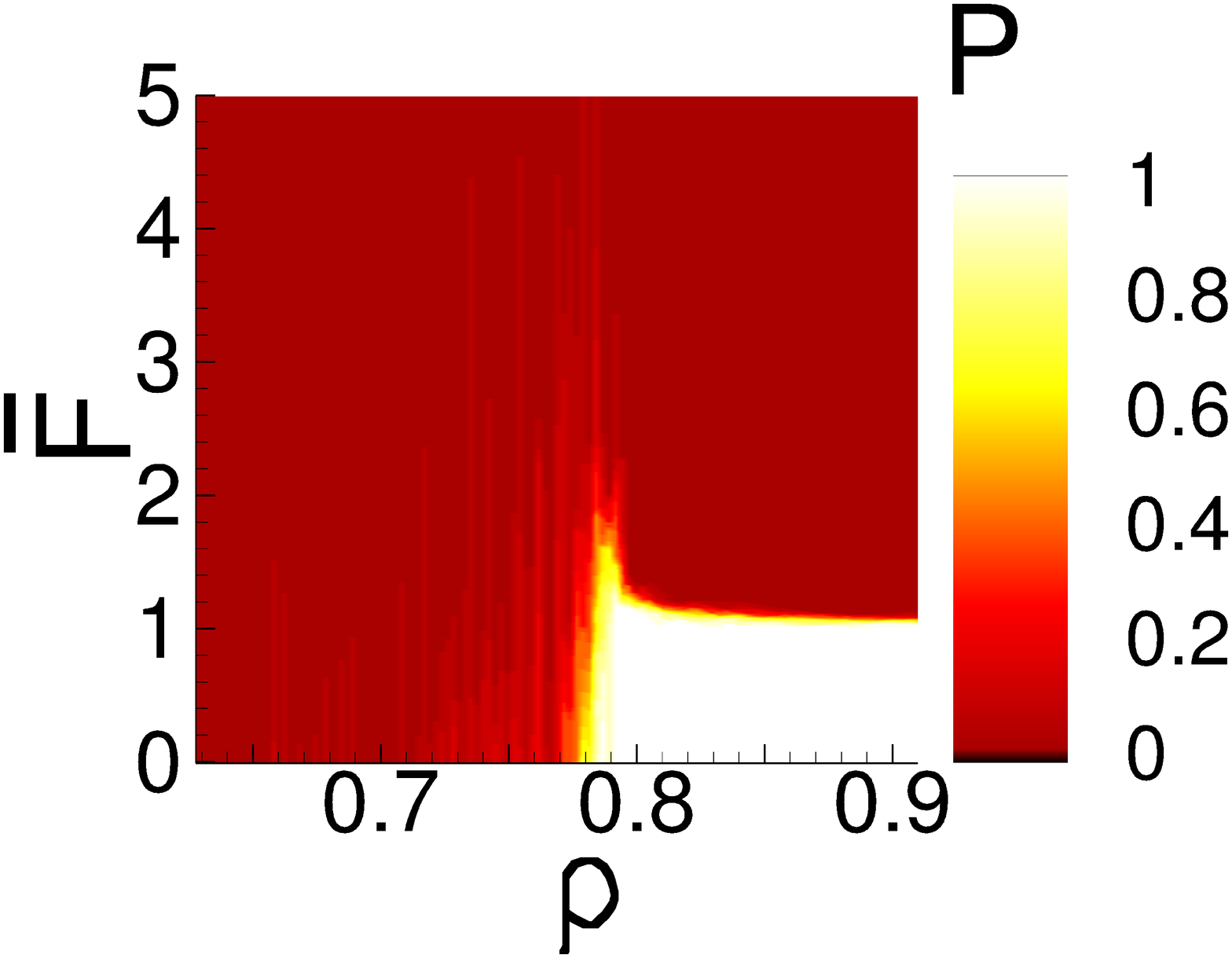,width=1.7in,height=1.3in}}
\subfloat[The percolation force threshold, $\bar F_p$, and the coordination number, $Z$.]{\epsfig{file=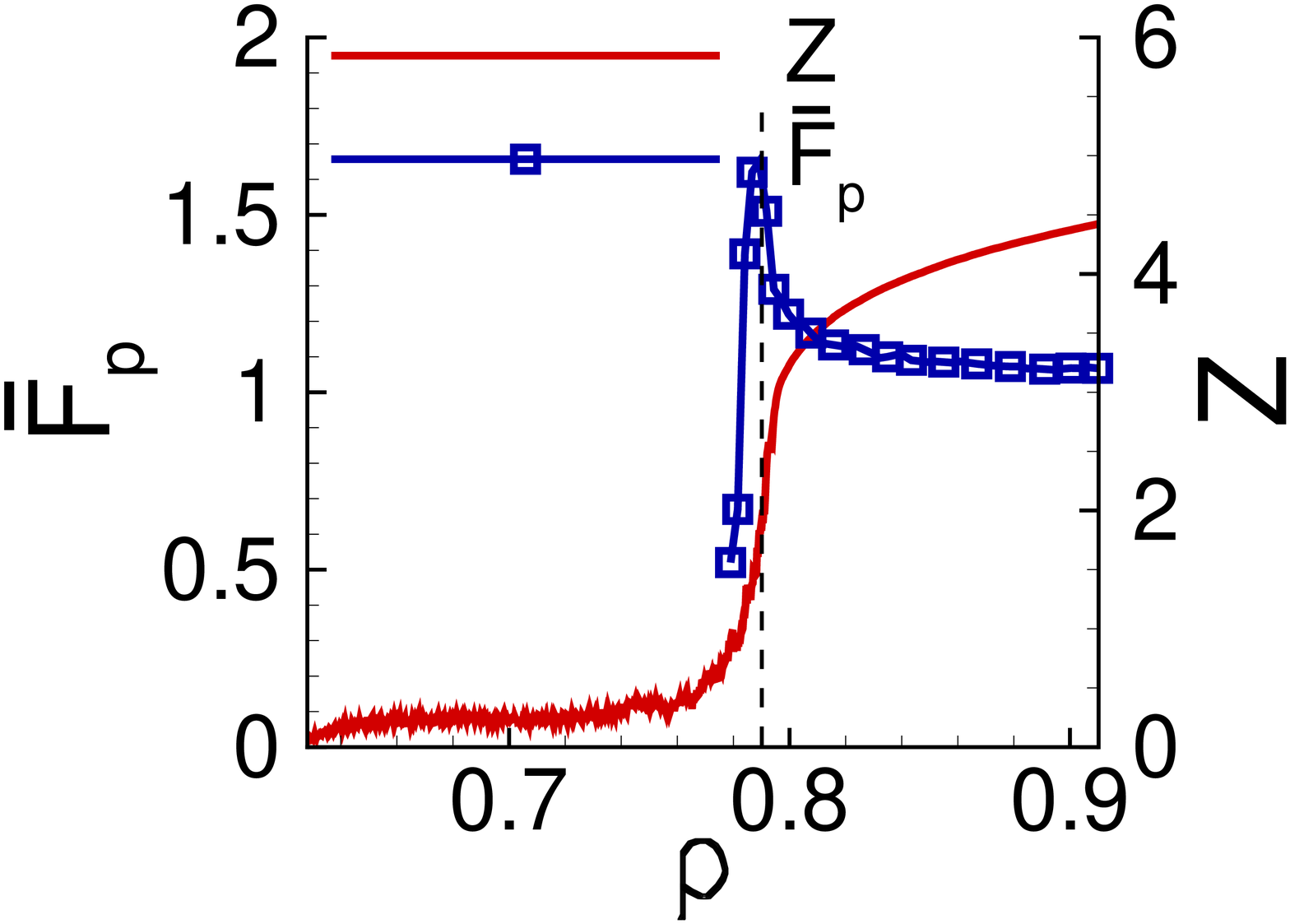,width=1.7in,height=1.3in}}
\caption{ (Color online) Reference system, averaged over $20$ realizations.}
\label{Pc_refcase}
\end{figure}
 As an example, 
Fig.~\ref{part_refcase}(b) shows the same system
as in Fig.~\ref{part_refcase}(a) with force threshold $\bar F =1$.   While the system shown in Fig.~\ref{part_refcase}(a) clearly 
percolates (contact percolation), it is not immediately obvious whether the system shown in Fig.~\ref{part_refcase}(b) does.

In describing percolation properties, we use the following quantities, all based on averaging over multiple
realizations: $P(\rho,\bar F)$, the percolation probability; $\bar F_p$,  the percolation force threshold, defined by $P(\rho,\bar F_p)=0.5$; 
and $P_c (\rho)$, the contact percolation probability, defined as $P_c (\rho) = P(\rho, 0)$.   In addition, we will use $Z$, the coordination
number, measuring average number of contacts per particle; a sharp increase of the $Z$ curve is typically associated with the jamming 
transition, see, e.g.~\cite{majmudar07a}. We note that the listed quantities also depend on the number of particles, $N$, and on the compression 
speed, $v_c$; this dependence will be discussed later in the paper.  For the simplicity of notation, we do not include this dependence 
explicitly in the notation. 

Figure~\ref{Pc_refcase}(a) shows  $P(\rho,\bar F)$,  
for the  reference system.  We see that, starting at  $\rho\approx0.77$, there is a 
percolation transition; note that if we vary $\bar F$ and keep $\rho$ fixed, this transition is rather sharp for large $\rho$'s and 
more spread out for  $\rho \in [0.77,0.81]$.  
To describe various transitions that take place as the system is compressed, we define: 
$\rho_J$, at which jamming, defined here as the $\rho$ at which the $Z$ curve has an inflection point, takes place
(later in the text we also show that  at $\rho_J$ rapid increase in pressure (measured at the domain boundaries) occurs, supporting this definition of $\rho_J$);  
and $\rho_p$,  at which contact percolation, defined as $P_c(\rho_p)=0.5$ occurs. 
Figure~\ref{Pc_refcase}(b) shows $Z$ and $\bar F_p$; we find from the data shown that $\rho_J \approx 0.79$ (the vertical dashed line in the figure).  
Note that just below $\rho_J$, there is a strong force network that percolates, as shown by large $\bar F_p$.
The dominant maximum of $\bar F_p$ calls for consideration of another transitional $\rho$ at which this maximum 
occurs: however, we find that this transition is always sandwiched between $\rho_p$ and $\rho_J$, so we will not discuss
it in more details here. 
\begin{figure}[ht!]
\begin{center}
\subfloat[$v_c = v_0$.]{\hspace{-0.5cm}\epsfig{file=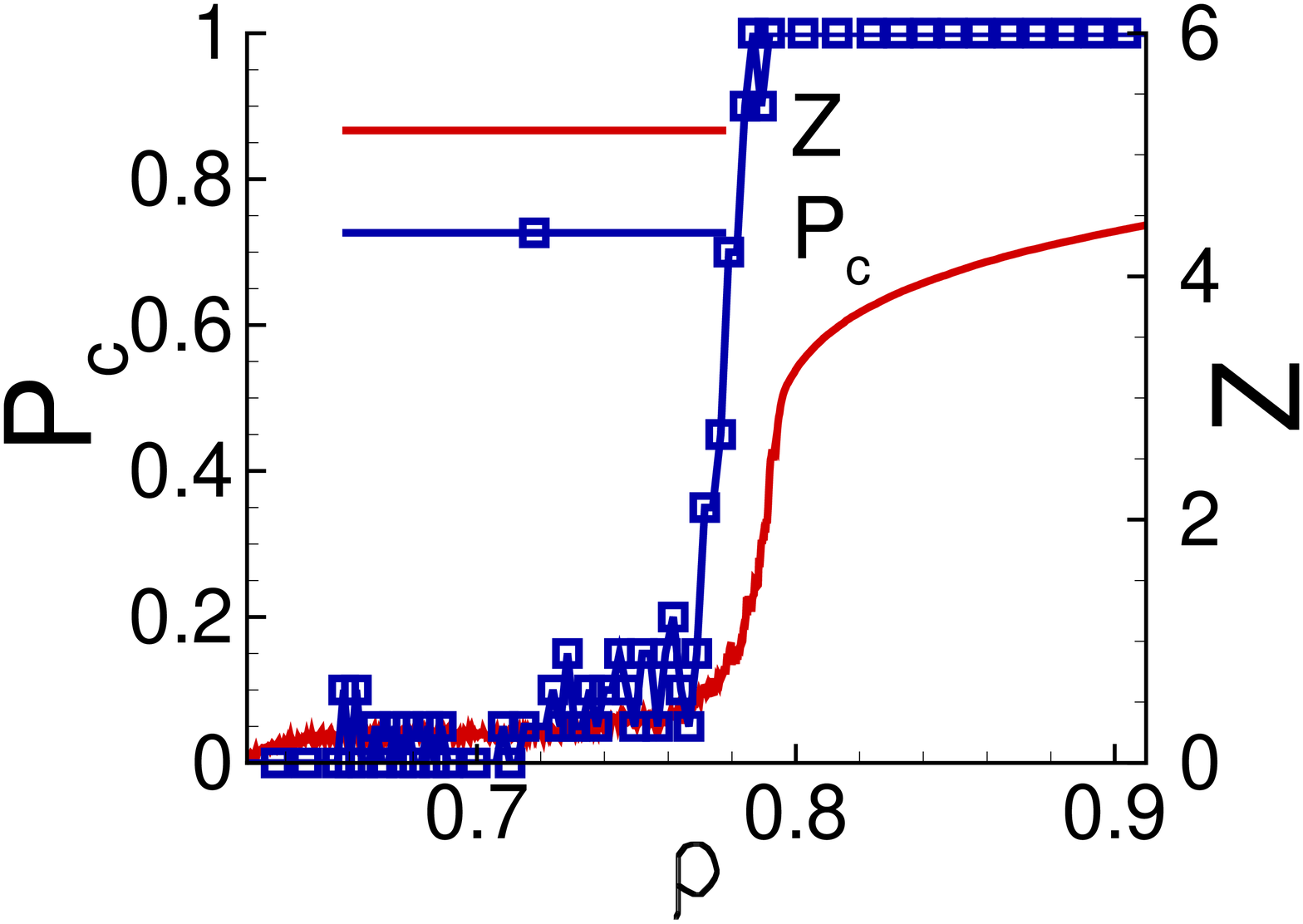,width=1.7in,height=1.3in}}
\subfloat[$v_c = 0$.]{\epsfig{file=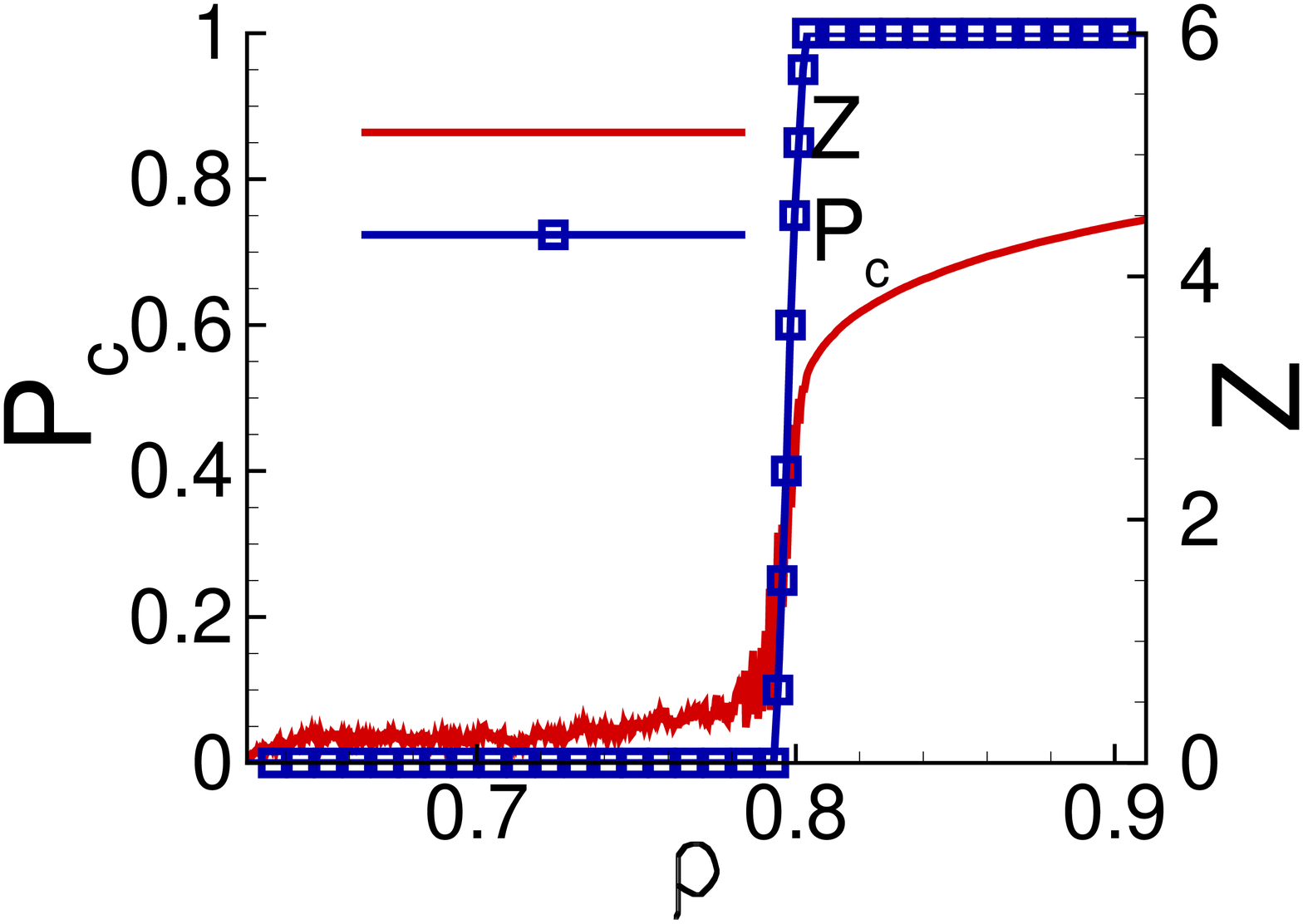,width=1.7in,height=1.3in}}
\end{center}
\caption{(Color online) Reference system: the percolation probability, $P_c$, and $Z$.}
\label{Pc_F_eq_0_with_Z_refcase}
\end{figure}
 
Figure~\ref{Pc_F_eq_0_with_Z_refcase}(a) shows $Z$  and  $P_c$ for the reference system.  While there is some noise in the results, 
one can still obtain an accurate value for $\rho_p \approx 0.776$. [For this, and all other results involving $\rho_p$ and $\rho_J$, uncertainty of the
\begin{figure}[ht!]
 \subfloat[fixed compression rate]{\epsfig{file=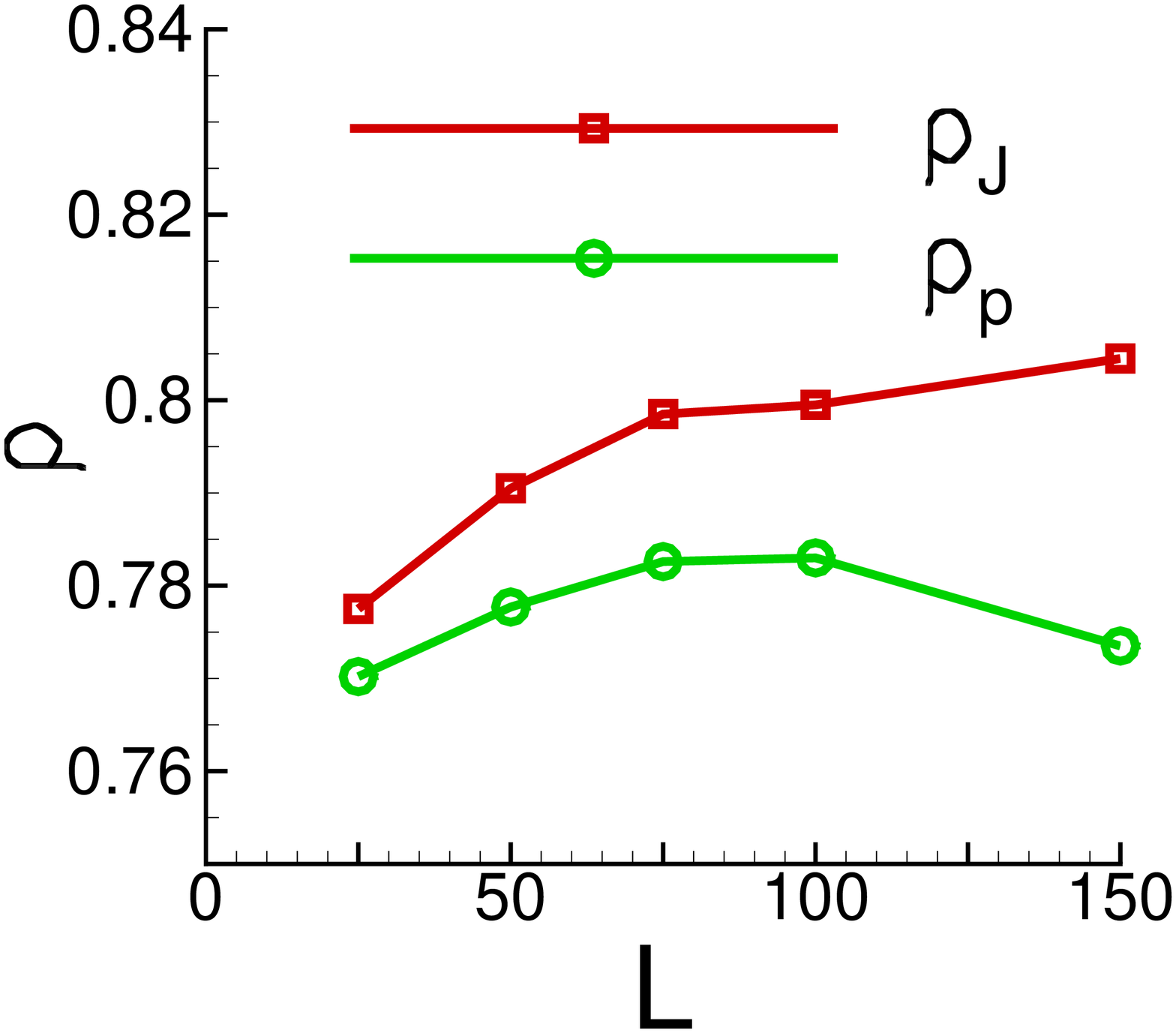, width=1.7in}}
 \subfloat[fixed compression speed]{\epsfig{file=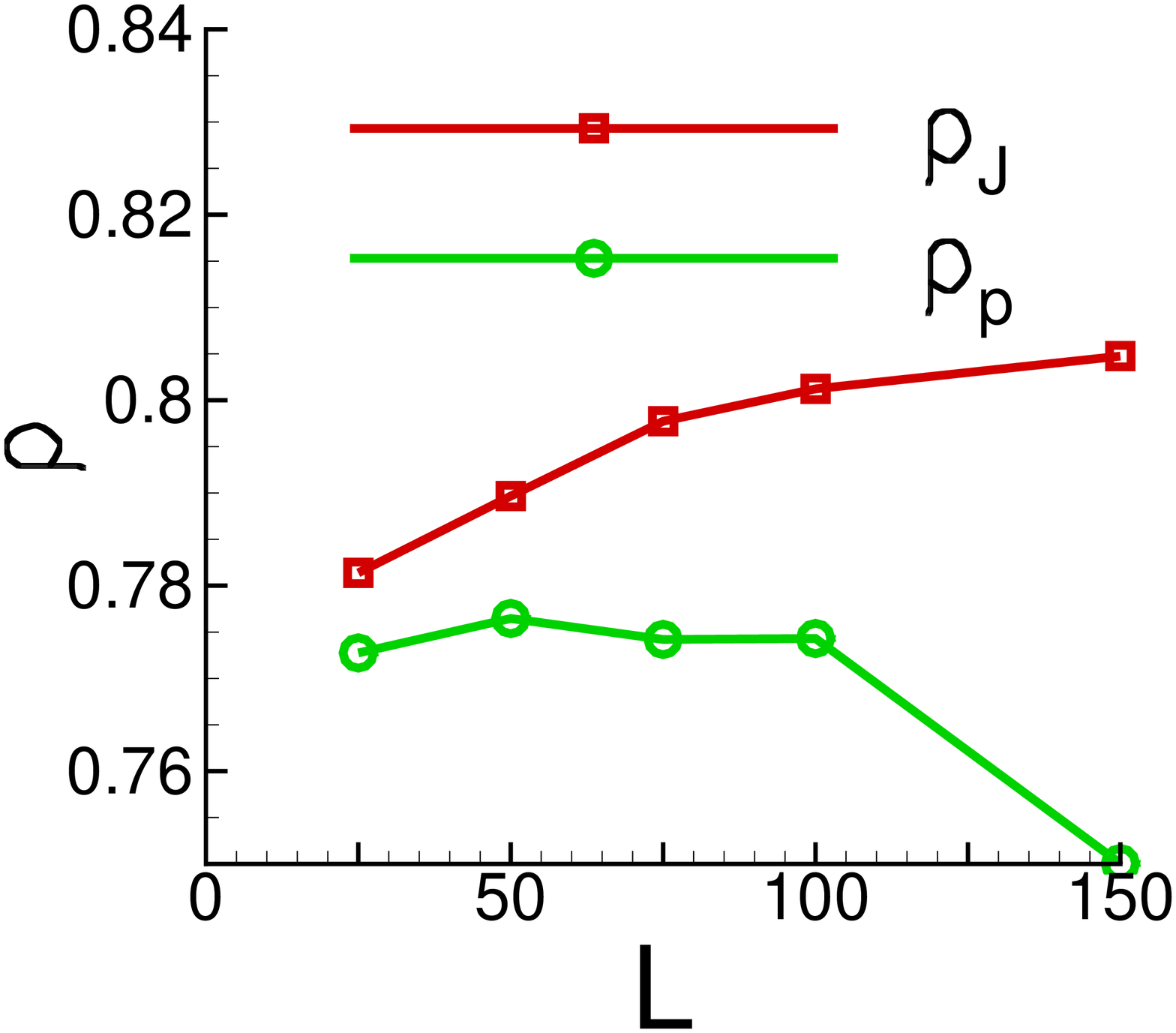,width=1.7in}}
 \caption{(Color online) Influence of system size on $\rho_p$ and $\rho_J$ for fixed compression rate and for fixed compression speed.}\label{rhop_rhoj_vs_L}
 \end{figure}
results is such that the results are accurate up to three significant digits: for $\rho_J$ we use standard error to estimate uncertainty, and 
for $\rho_p$ we estimate the range over which  $0.4\leq P_c(\rho)\leq 0.6$.]
Therefore, the results for our reference system suggest that 
$\rho_p < \rho_J$, and the question is whether this finding is robust with respect to the changes of the system parameters and of the protocol used.  Before proceeding, we note that although there are some differences between realizations, for all of
them we find consistently (for the considered system) that $\rho_p$ and $\rho_J$ differ by a nonvanishing amount.

Regarding the system parameters, we start by discussing the influence of polydispersity, measured by $r_p$,  and friction coefficient, $\mu$. Table~\ref{tab:parameters} shows
the results for $\rho_p$ and $\rho_J$, and we observe that both $\rho_p$ and $\rho_J$ are monotonously decreasing functions of these
two parameters; in particular  the results for $\rho_J$ are consistent with the ones from literature (see~\cite{epl12} and the references therein).   The finding that is perhaps
more relevant for the present discussion is that the difference between $\rho_p$ and $\rho_J$ remains as $r_p$ and $\mu$ are varied.

Next  we discuss the influence of system size; note that this issue has been discussed extensively in the context of random percolation (see e.g.~\cite{stauffer}).
Here, the context is more complicated since the system considered is dynamic, and one has to decide on coupling of relevant spatial 
and temporal scales.  We have considered two scenarios for the systems of different size: one where the rate of the change of $\rho$ is kept constant, 
\begin{table}[ht!]
\begin{center}
 \begin{tabular}{|l||c|c|c|c|c|r|}
  \hline
  \multicolumn{7}{|c|}{$\mu$}\\
  \hline
   &0.0&0.1&0.2&0.3&0.4&0.5\\
  \hline
  $\rho_J$&0.827&0.812&0.802&0.797&0.796&0.789\\
  \hline
  $\rho_p$&0.815&0.799&0.792&0.784&0.781&0.776 \ \\
  \hline
  \end{tabular}
  \begin{tabular}{|l||c|c|c|c|r|}
    \hline
    \multicolumn{6}{|c|}{$r_p$}\\
    \hline
     &0.0&0.1&0.2&0.3&0.4\\
    \hline
    $\rho_J \ $&0.804 \ \ &0.797 \ \ &0.789 \ \ &0.7834&0.782 \\
    \hline
    $\rho_p \ $&0.786 \ \ &0.784 \ \ &0.776 \ \ &0.771 \ \ &0.766 \\
    \hline
  \end{tabular}
 \begin{tabular}{|l||c|c|c|c|r|}
  \hline
  \multicolumn{6}{|c|}{$v_c/v_0$}\\
  \hline
     &0.0&0.02&0.05&0.1&1.0\\
  \hline
  $\rho_J$&0.798 \ \ &0.799 \ \ &0.798 \ \ &0.792 \ \ &0.789 \ \\
  \hline
  $\rho_p$&0.798 \ &0.794 \ &0.791 \ &0.786 \ &0.776 \ \\
  \hline
 \end{tabular}
  \caption{Influence of $\mu$, $r_p$ and $v_c$ on $\rho_p$ and $\rho_J$ for a continuously compressed system (the parameters not specified correspond to the reference case). }
    \label{tab:parameters}
\end{center}
\end{table}
and the one where the  compression speed ($v_c$) is fixed. While the details of the results vary depending on the choice of the 
scenario, we find that the difference between $\rho_p$ and $\rho_J$ remains  non-zero (and typically increases as a function of $L$) 
for the both scenarios and for the system sizes defined by $L = 50,~75,~100,~150$: Figure~\ref{rhop_rhoj_vs_L} shows the dependence of $\rho_p$ and $\rho_J$ behavior on the system size using two aforementioned protocols. Figure~\ref{rhop_rhoj_vs_L}(a) shows results for the fixed compression rate; the compression velocity, $v_c$, is increased with $L$ so that the rate $v_c/L$ is constant. Figure~\ref{rhop_rhoj_vs_L}(b) shows $\rho_p, \rho_J$ when we keep $v_c$ constant as $L$ increases. For both protocols -- fixed compression rate and speed -- we observe increased difference between $\rho_p$ and $\rho_J$ as $L$ is increased. 

Since the reference system is exposed to  a nonvanishing compression rate, there is also the question of rate-dependence, as already alluded above.  
To explore this issue,  we carry out simulations with progressively smaller speed of compression, using 
$v_c= v_0/10,v_0/20$ and $v_0/50$.  We find that the $P_c$ transition
becomes sharper as $v_c$ decreases, indicating that $\rho_p$ is affected by $v_c$; in general,  for a fixed $\rho$,  the particles are less likely 
to percolate for smaller $v_c$ and therefore $\rho_p$ increases as $v_c$ decreases.   
Both $\rho_p$ and $\rho_J$ are shown in 
Table~\ref{tab:parameters}.  While both $\rho$'s increase as $v_c$ decreases, 
the crucial finding is that the difference between them becomes smaller for slower compression.    
The question remains whether $\rho_p$ and $\rho_J$ collapse to a single value in the limit $v_c\rightarrow 0$.  
To answer this, we consider a modified protocol such that we interject relaxation steps in our compression (we reference this protocol
by $v_c  =0$).   More precisely, 
after compressing the system by $\delta \rho = 0.001$, we check whether there is a percolating cluster.  If not,  we proceed with compression; if yes, the system is relaxed until 
percolation disappears,  and then the system is further compressed. 
We carry out this procedure until such $\rho_p$ that percolating cluster does not disappear after relaxation
(for all considered simulations, the system always percolates above $\rho_p$ found using relaxation protocol, or in other words, percolation is never found to disappear as 
a system is further compressed).  
Figure~\ref{Pc_F_eq_0_with_Z_refcase}(b) shows $P_c$ and $Z$  for the relaxed system, suggesting much smoother and sharper evolution of $P_c$ through $\rho_p$.  
Table~\ref{tab:parameters} shows that for the reference system and $v_c = 0$, $\rho_p$ and $\rho_J$ collapse to the same point, within the 
available accuracy. We have reached the same finding for the other systems listed in Table~\ref{tab:parameters}, 
including monodisperse frictionless system  - while this particular system is known to show
different behavior due to partial crystallization~\cite{epl12}, it still leads to $\rho_p = \rho_J$. We have also verified that the finding 
$\rho_p=\rho_J$ still holds when different system sizes are considered.  

This finding of collapse of percolation and jamming transitions appears to be different from the one in \cite{ohern_pre_12}, where it was found that $\rho_p$ and $\rho_J$
differ.   The source of the difference seems to be the use of overdamped dynamics in~\cite{ohern_pre_12}; this effect apparently keeps the particles together and leads to percolation even for small $\rho$'s.   
\begin{figure}[ht!]
 \epsfig{file=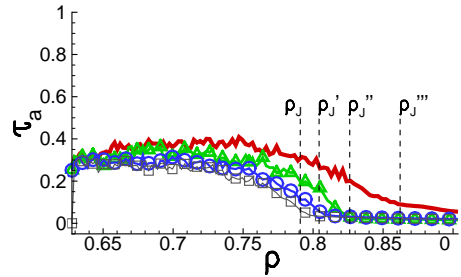,width=2.7in}
 \caption{(Color online) Anisotropy of the stress tensor of $r_p=0.2,\mu=0.5$ (squares), $r_p=0.0,\mu=0.5$ (circles), $r_p=0.2,\mu=0.0$ (triangles) and $r_p=0.0,\mu=0.0$ (thick line) systems as a function of packing fraction, $\rho$. 
 Respective jamming transitions, $\rho_J, \rho_J', \rho_J'', \rho_J'''$, are depicted by a dashed line.}\label{anisotropy}
\end{figure}
We find, however, that, within the particle interaction model considered in the present paper, based on (constant) coefficient of restitution, $e_n$, the finding $\rho_p = \rho_J$ persists even for very small $e_n \approx 0$, suggesting that the finding reported here is robust, within the framework of the implemented particle interaction model.  

While the findings obtained in quasi-static limit are of main interest, one should note that in the context of particulate matter,  percolation and jamming transitions typically
involve dynamics, even if very slow one.   Close to $\rho_J$, the relevant time scales diverge in the limit of infinite system size, and therefore, one could expect that for any 
sufficiently large system, even very slow dynamics may lead to (arbitrarily small) differences between $\rho_p$ and $\rho_J$.    Therefore, it should not be surprising 
if differences are found between $\rho_p$ and $\rho_J$ for slowly evolving spatially extended particulate systems.

To close our discussion focusing on repulsive systems, we discuss whether the implemented compression protocol may induce an anisotropy, 
possibly influencing the results.   For this purpose, we compute the stress tensor and the distribution of the angles of contact between the particles.
 For brevity, we consider here only the compression by $v_0$.  The stress anisotropy, $\tau_a$, is defined by
 \begin{equation}
\tau_a={\sigma_1-\sigma_2\over\sigma_1+\sigma_2}
\end{equation}
with $\sigma_1, \sigma_2$ the principal eigenvalues of the Cauchy stress tensor $\sigma$, specified by 
$\sigma_{ij}=1/(2A)\sum_{c_k, p}(F_ir_j+F_jr_i)$ as a sum over all inter-particle contacts $c_k$ for all particles $p$; 
\begin{figure}[ht]
\subfloat[$\rho=0.73$ ]{ \epsfig{file=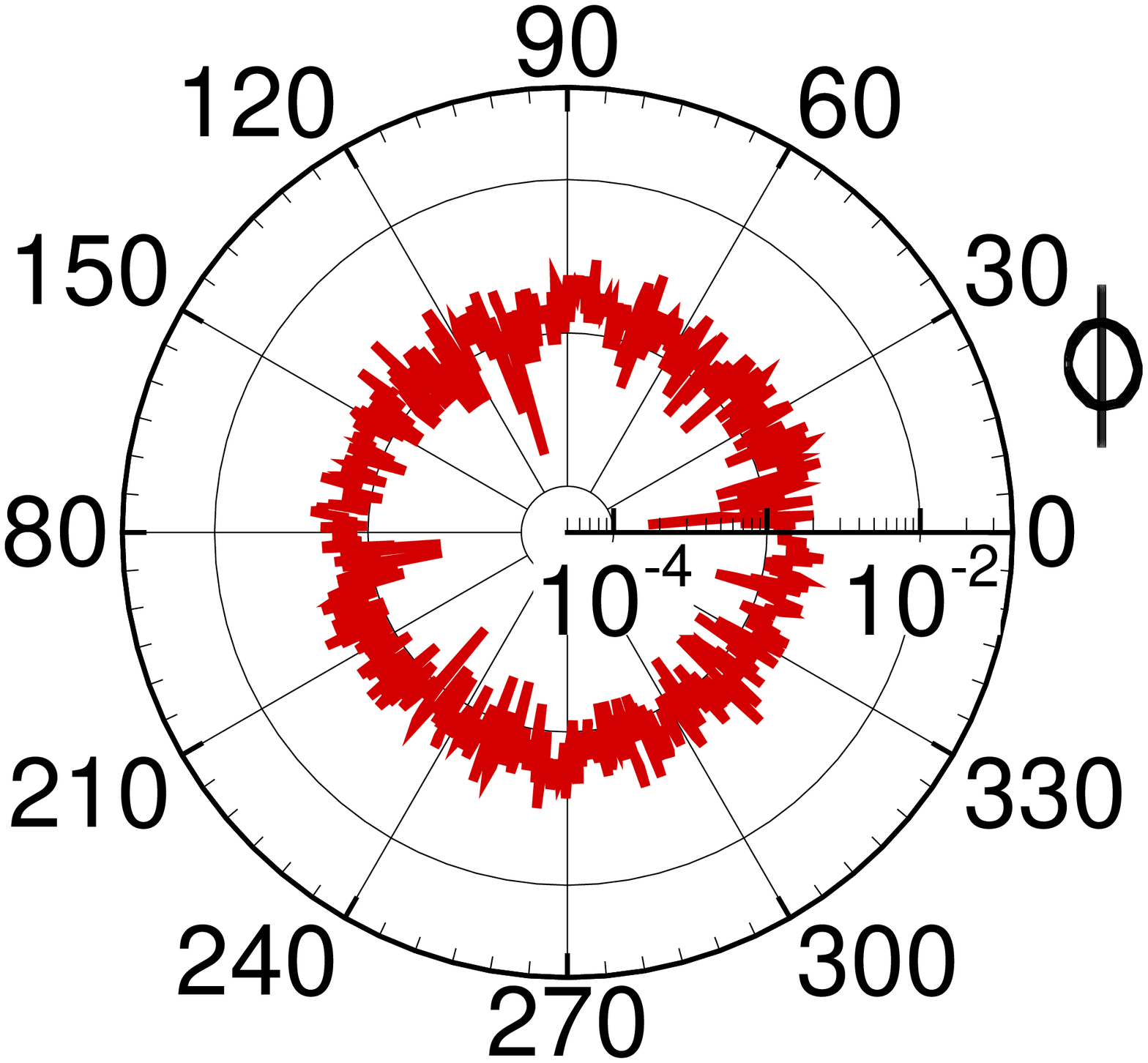,width=1.7in}}
\subfloat[$\rho=0.90$ ]{ \epsfig{file=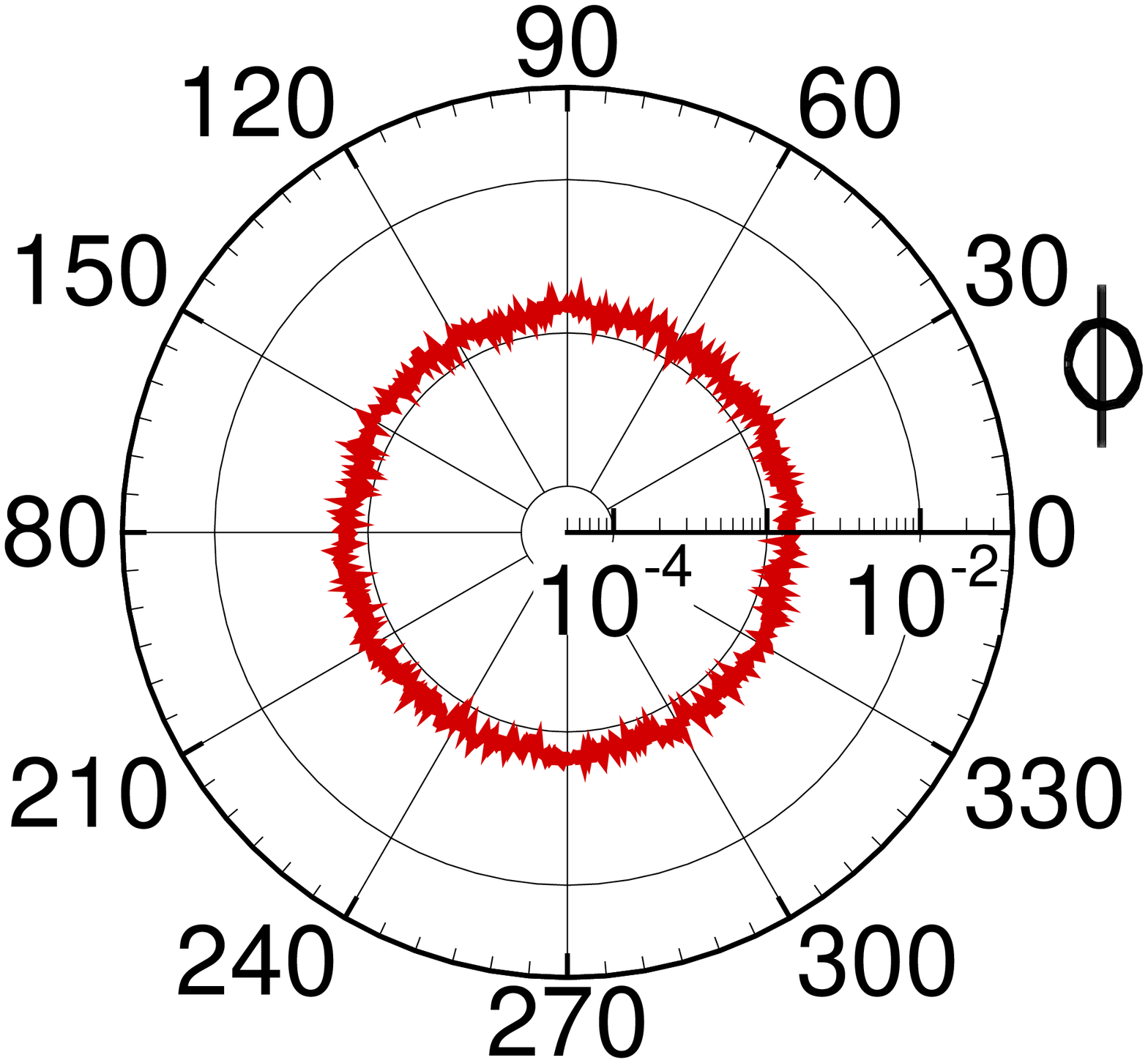,width=1.7in}}\\
\subfloat[$\rho=0.73$ ]{ \epsfig{file=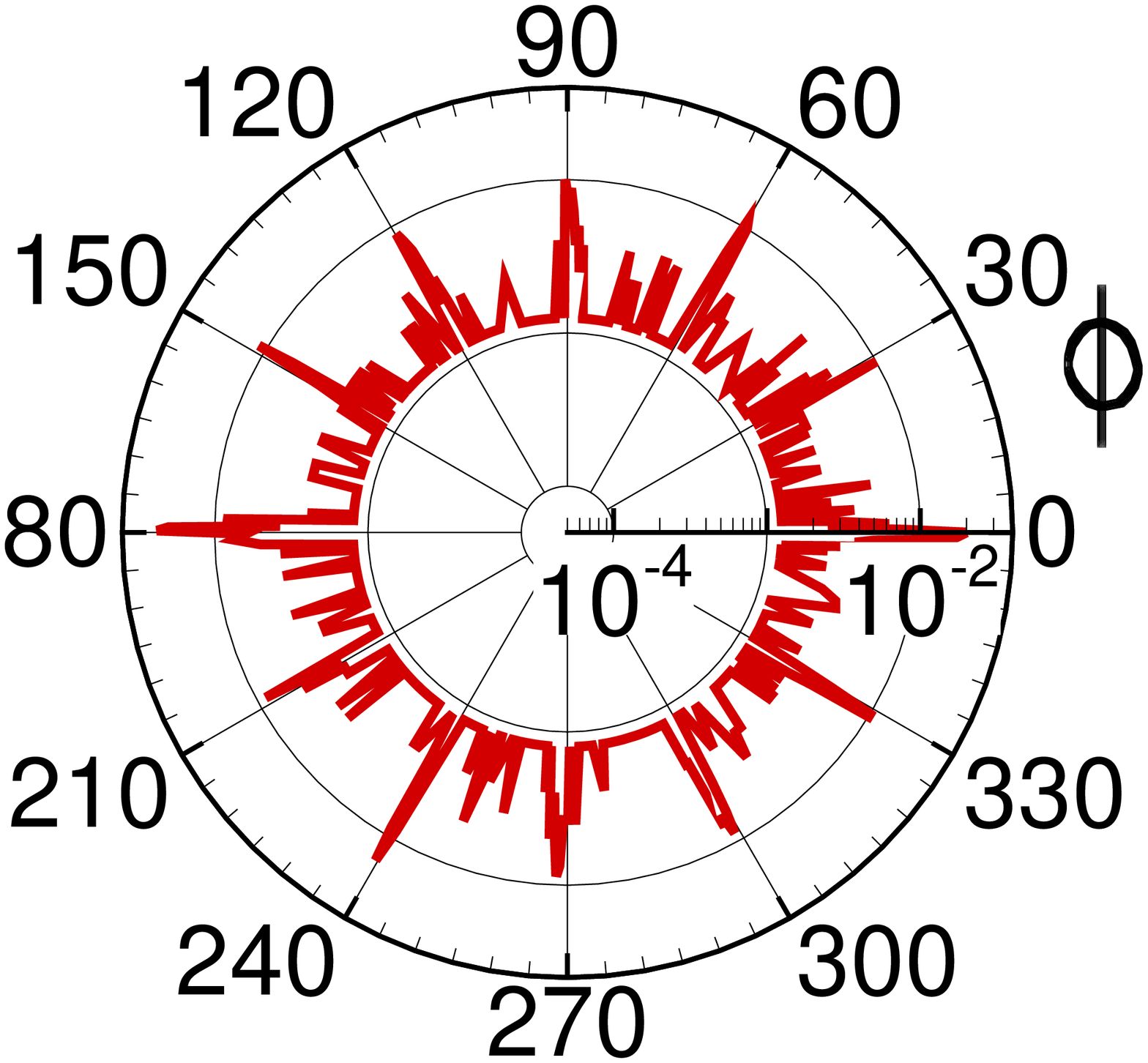,width=1.7in}}
 \subfloat[$\rho=0.90$ ]{ \epsfig{file=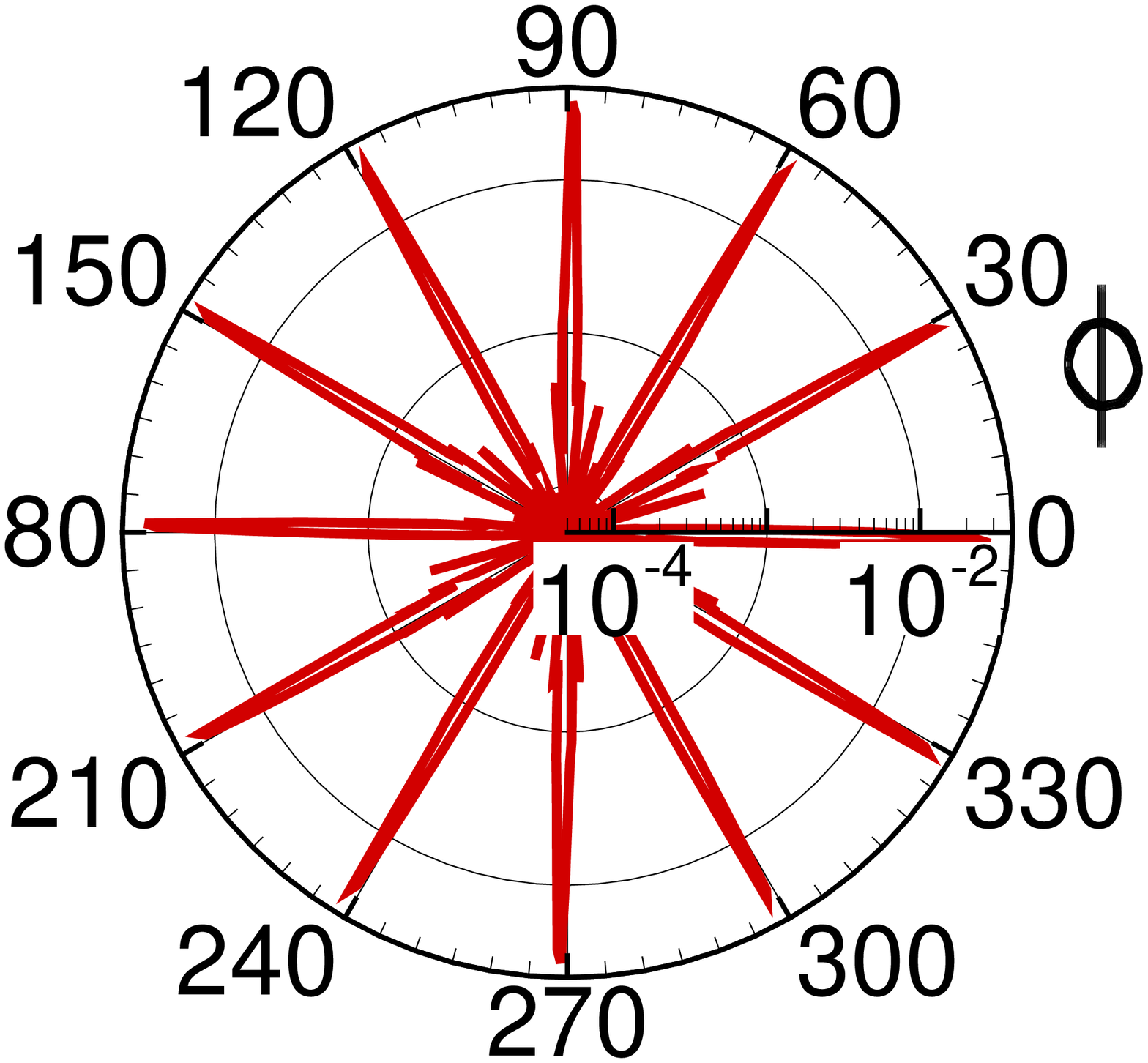,width=1.7in}}
 \caption{(Color online) Distribution of the angles of contacts for the reference (a), (b) and $r_p=0.0,~\mu=0.0$ system (c), (d).  In these polar 
 plots, the azimuthal coordinate, $\phi$, corresponds to the angle between the line connecting the centers of contacting particles 
 and the +$x$ axis,  and the radial one to the probability of observing given $\phi$.}\label{angle_dist}
\end{figure}
(wall particles as well as the contacts of interior particles with the wall particles are not included here). 
Here, $A$ is the total area of the system, $r_i,~r_j$ are the $x$ and $y$ components of the vector pointing from the center of particle $p$ 
towards the particle contact $c_k$. $F_i,~F_j$ denote the $x$, $y$ components of the interparticle force at the contact $c_k$.

Figure~\ref{anisotropy} shows $\tau_a$ as a function of $\rho$. We depict jamming transitions, $\rho_J, \rho_J', \rho_J''$ and $\rho_J'''$ by  dashed lines for $\mu=0.5, r_p=0.2$ (reference system), $\mu=0.5, r_p=0.0$, $\mu=0.0, r_p=0.2$ and $\mu=0.0, r_p=0.0$, respectively.   
\begin{figure}[ht]
\begin{center}
 \subfloat[$P_c$ and $Z$.]{\hspace{-0.5cm}\epsfig{file=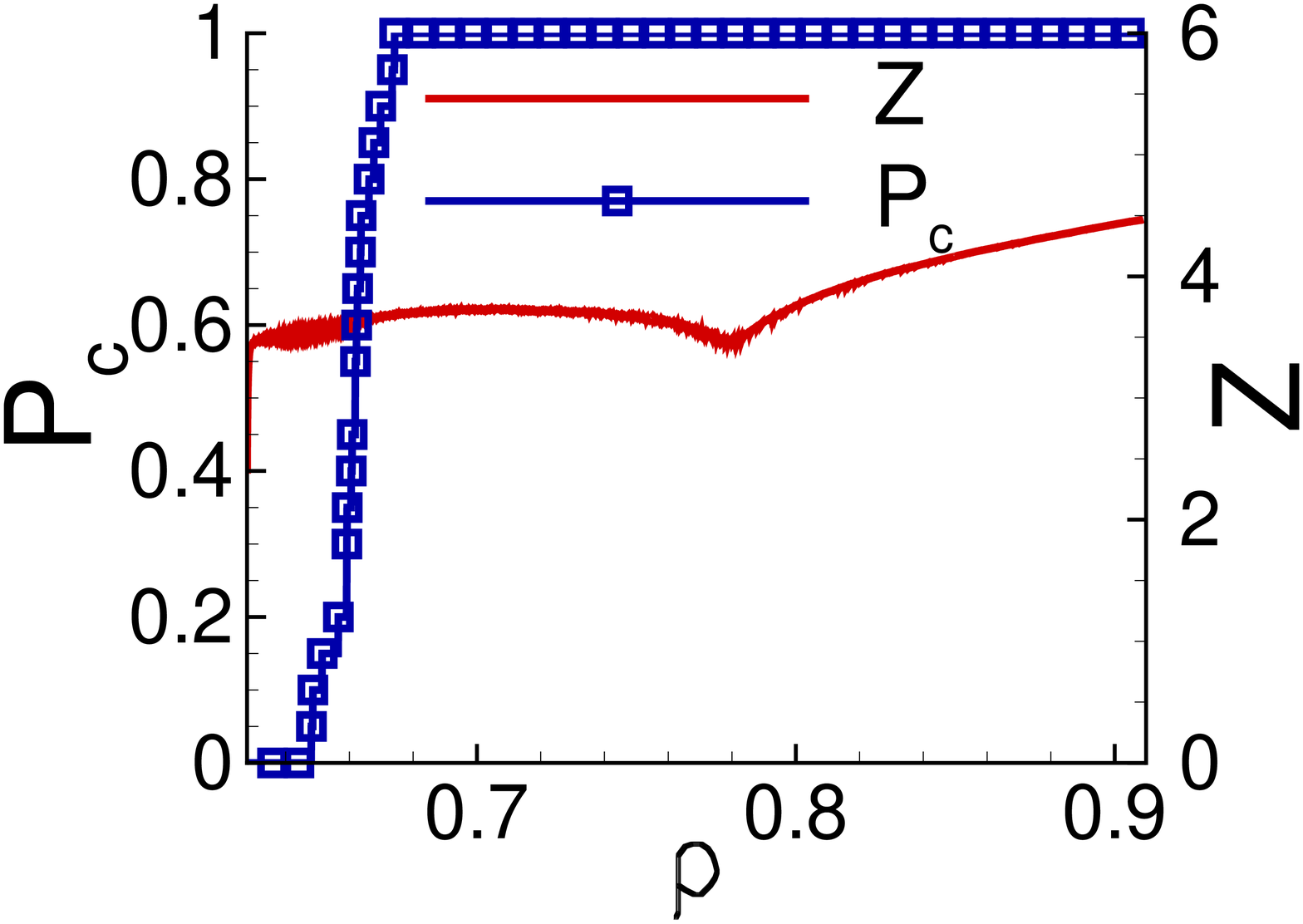,width=1.7in}}
 \subfloat[Number of particles with $cn=2-6$ contacts (circle, square, delta, gradient and thick line, consecutively).]{\epsfig{file=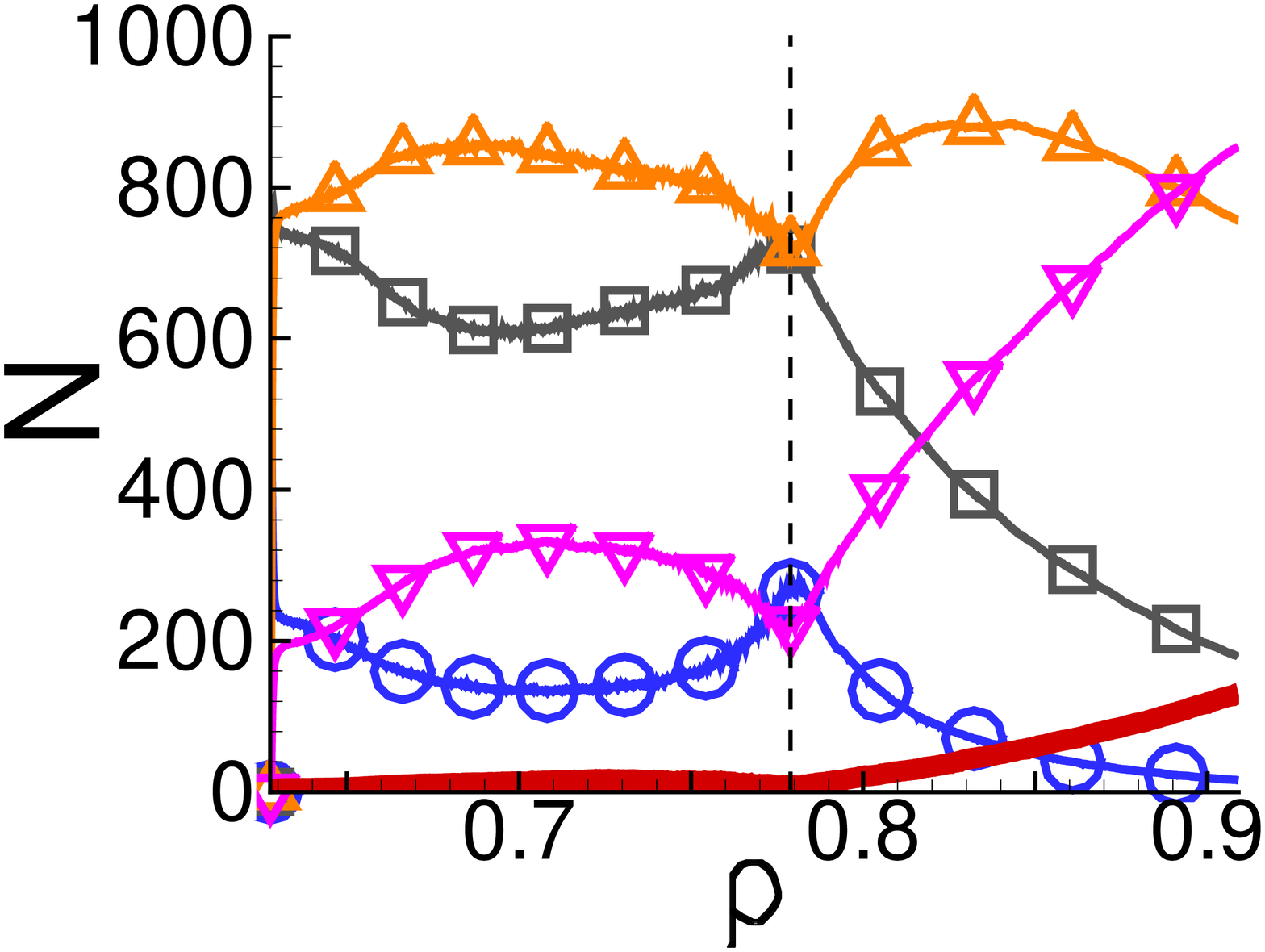,width=1.7in,height=1.35in}}\\
 \subfloat[Reference relaxed system.]{\epsfig{file=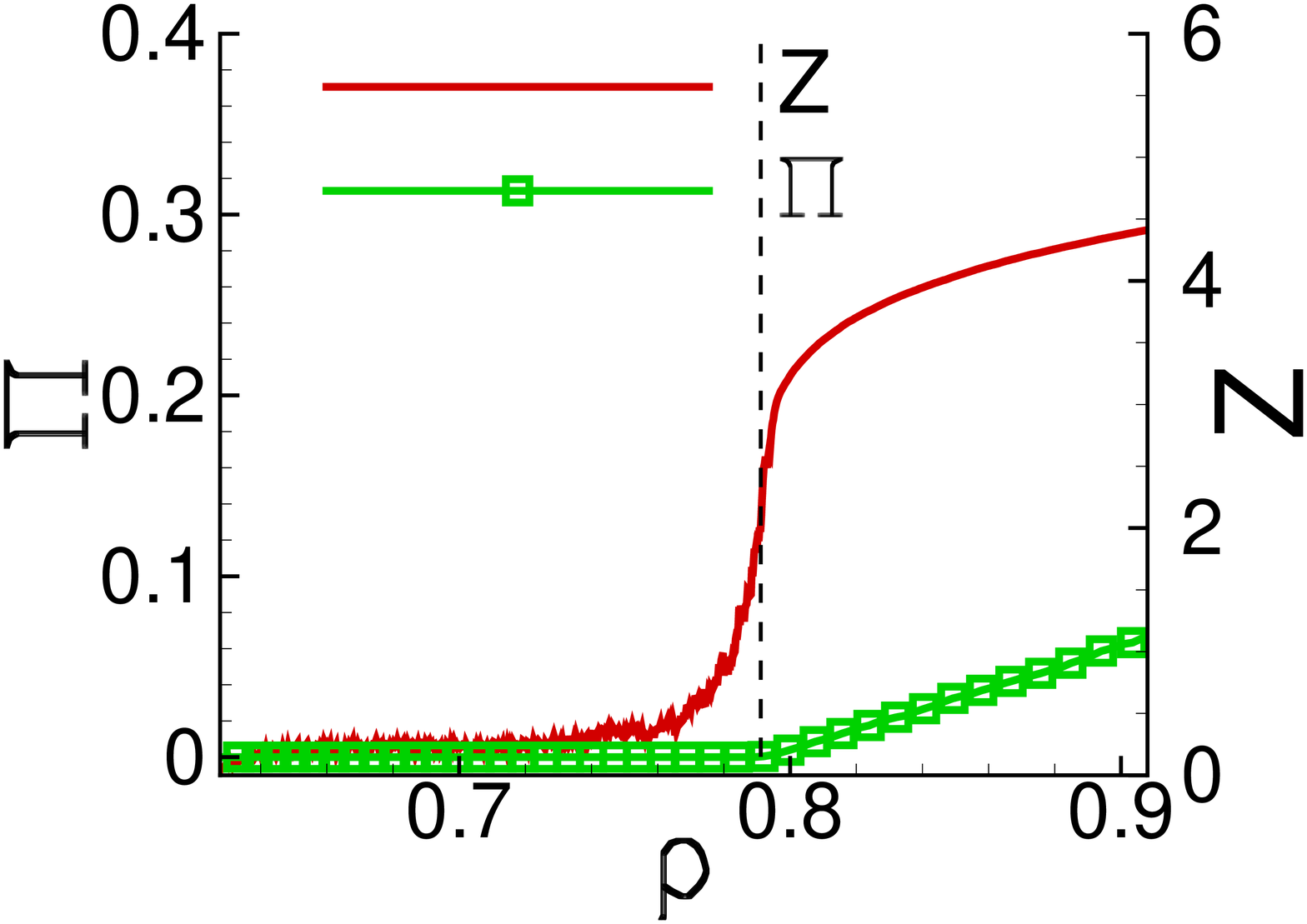,width=1.7in}}
 \subfloat[Cohesive relaxed system.]{\epsfig{file=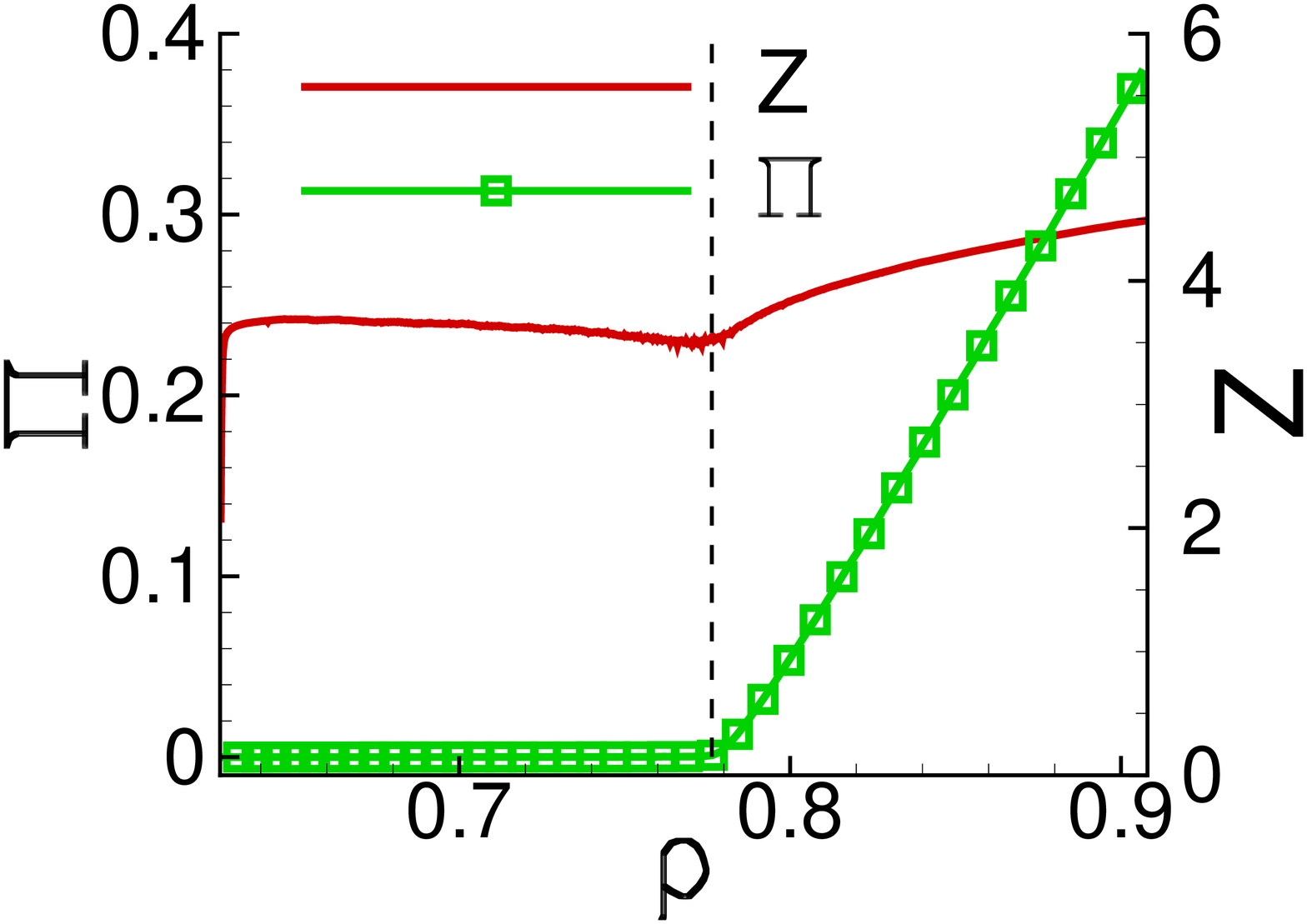,width=1.7in}}
 \caption{ (Color online) (a), (b) Cohesive relaxed system. (c), (d) Pressure on the walls, $\Pi$, and $Z$.   Dashed lines correspond 
 to $\rho_J$ (and to $\rho_p$ in (c)); in (a), $\rho_p$ is shown by dotted line.
 }\label{fig:Pc_Z_wet_relaxed_wallP}
 \end{center}
\end{figure}
While far below the jamming (and percolation) transitions, the anisotropy measured by $\tau_a$ may be present, close to $\rho_p$ and $\rho_J$, $\tau_a \ll 1$ for all systems considered, 
showing that the systems are essentially isotropic for the packing fractions of relevance here.
Above jamming points, $\tau_a$ is even smaller. 

Figure~\ref{angle_dist} shows the distribution of contact angle, $\phi$, for the reference system, the parts (a), (b), and for the $\mu=0.0$, $r_p =0.0$ 
system, the parts (c), (d).   Most importantly, this figure shows symmetric distribution of $\phi$'s.     In addition, by comparing the results of the reference case 
with the ones obtained for monodisperse frictionless, we also observe the influence of partial crystallization on the latter, for large packing fractions.

\subsection{Cohesive systems}
\label{sec:cohesive}

Here, we discuss the effect of cohesion on percolation and jamming. We have considered few different `strengths' of cohesion (specified by the distance, $s_c$), at
which capillary bridges break; for brevity here we present results only for `weak' cohesion, specified by small distance at which 
capillary bridges break, $s_c \approx 0.0028\ll 1$ (see Sec.\ref{sec:sim}).  We focus on the relaxed reference system.
Figure~\ref{fig:Pc_Z_wet_relaxed_wallP}(a) shows  that the percolation transition occurs very close 
to (the starting value) $\rho = 0.63$.  The  $Z$ curve remains at high values for all considered $\rho$'s, but we note that there is a
kink in the $Z$ curve at $\rho\approx 0.783$. The kink and consecutive increase of $Z$ suggest 
that the system undergoes a transition. To verify that this transition corresponds to $\rho_J$, we consider the pressure on the system walls, $\Pi$.  
Figure~\ref{fig:Pc_Z_wet_relaxed_wallP}(c and d) shows this pressure (force/length, in dimensionless units) for both the reference system, and for the cohesive one.
We see that for the reference system an increase of $\Pi$ occurs at $\rho_J$  (inflection point of the $Z$ curve). Figure~\ref{fig:Pc_Z_wet_relaxed_wallP}(d) shows that 
an increase in $\Pi$ and the kink in the $Z$ curve occur at the same $\rho = \rho_J = 0.783$.     

Clearly, the difference between $\rho_J$ and $\rho_p$ is  significant for the considered cohesive system, consistently with the earlier 
work~\cite{lois}.     As expected, we find similar results for the systems characterized by larger $s_c$ (results not shown for brevity).  
The strong influence of weak cohesion on the $\rho_p$ and $\rho_J$ suggests that for any non-vanishing cohesion, one would find
differences between $\rho_P$ and $\rho_J$, with this differences disappearing only in the limit of $s_c \rightarrow 0$. As soon as 
there is no attractive force, the difference between $\rho_p$ and $\rho_J$ vanishes even in the limit of inelastic collisions, $e_n\rightarrow 0$.

One may ask about the origin of the `kink' in the $Z$ curves for the cohesive system. An intuitive explanation 
is as follows:  as compression starts, the particles immediately get in contact, form mini-clusters (consisting of a small number of particles), leading to rather large $Z$; due to the presence of cohesive forces, relaxation does not lead to breakup of the existing contacts. Therefore, as long as $\rho$ is small,  the mini-clusters do not break;  as $\rho$ grows, however, collisions start separating particles, leading to breakup of the mini-clusters and decreasing $Z$. At some point, when $\rho$ becomes sufficiently large so that all particles are effectively in contact, $Z$ starts growing again, and at the same $\rho$, $\Pi$ starts increasing.  To support this description, Fig.~\ref{fig:Pc_Z_wet_relaxed_wallP}(b) shows the number of particles ($N$) with $2,\dots, 6$ contacts ($cn$).  We observe that as $\rho_J$ is approached from below, the $cn=4,~5$ curves have
negative slope, suggesting breakup of the clusters (this breakup is presumably also partially responsible  for the positive slope of $c_n = 2,~3$ curves for the 
same values of $\rho$); at $\rho_J$ these trends reverse.

\section{Summary and conclusions}  
\label{sec:conclusions}

Percolation and jamming transitions of evolving particulate systems are non-trivial.   We find that these
transitions for repulsive particles interacting by a commonly used interaction model coincide for quasi-static systems;  this finding, together with the results
reported in~\cite{ohern_pre_12}, where these transitions are found to differ for particles following overdamped dynamics, suggests that the considered 
transitions may be influenced significantly by the type of interaction between the particles.    Furthermore, our 
finding is that any, even very slow dynamics may lead to the differences of the packing fractions at which percolation and jamming occur.   Therefore, in 
particular close to jamming, a careful exploration will be needed in order to distinguish the effects due to dynamics and due to, e.g., the type of interaction between
the particles.  In the same vein, we are also finding that even minor cohesive effects have a strong influence in particular on percolation transition.   

We hope that the present results will encourage carrying out careful experiments that will quantify further the predictions regarding the influence that 
dynamics, cohesion, and the nature of particle interaction have on percolation and jamming.    Our own research will continue in the direction of exploring the
effects of jamming and percolation in three spatial dimensions.  

{\it Acknowledgments}   This work was partially supported by the NSF Grant No. DMS-0835611. 

\bibliographystyle{apsrev}

\end{document}